\documentclass{article}

\usepackage{arxiv}

\usepackage[utf8]{inputenc} % allow utf-8 input
\usepackage[T1]{fontenc}    % use 8-bit T1 fonts
\usepackage[hidelinks]{hyperref}       % hyperlinks
\usepackage{url}            % simple URL typesetting
\usepackage{booktabs}       % professional-quality tables
\usepackage{amsfonts}       % blackboard math symbols
\usepackage{amsmath}
\usepackage{nicefrac}       % compact symbols for 1/2, etc.
\usepackage{microtype}      % microtypography
\usepackage{lipsum}		% Can be removed after putting your text content
\usepackage{graphicx}
\usepackage[authoryear, round]{natbib}
\usepackage{doi}

\title{Contributions of El Niño Southern Oscillation (ENSO) Diversity to Low-Frequency Changes in ENSO Variance}

\date{} 					% Or removing it

\author{
Jakob Schlör \\
Machine Learning in Climate Science \\
University of Tübingen, Germany\\
\texttt{jakob.schloer@uni-tuebingen.de} \\
\And
Felix Strnad \\
Machine Learning in Climate Science \\
University of Tübingen, Germany\\
\And
Antonietta Capotondi \\
Cooperative Institute for Research in Environmental Sciences \\
University of Colorado, Boulder, CO \\
NOAA/Physical Sciences Laboratory, Boulder, CO \\
\And
Bedartha Goswami \\
Machine Learning in Climate Science \\
University of Tübingen, Germany\\
}

\renewcommand{\headeright}{Schlör et al.}
\renewcommand{\undertitle}{}
\renewcommand{\shorttitle}{Contributions of ENSO Diversity to Low-Frequency Variability}

\hypersetup{
pdftitle={Contributions of ENSO Diversity to Low-Frequency Variability},
pdfsubject={preprint},
pdfauthor={Jakob Schlör, Felix Strnad, Antonietta Capotondi, Bedartha Goswami},
pdfkeywords={El Nino Southern Oscillation, Unsupervised clustering},
}

\begin{document}

% Word count: 3986
%TC:ignore 
%\detailtexcount{main_pcgmm} 
%\newpage
%TC:endignore
\maketitle

\begin{abstract}
El Niño Southern Oscillation (ENSO) diversity is characterized based on the longitudinal location of maximum sea surface temperature anomalies (SSTA) and amplitude in the tropical Pacific, as Central Pacific (CP) events are typically weaker than Eastern Pacific (EP) events. SSTA pattern and intensity undergo low-frequency modulations, affecting ENSO prediction skill and remote impacts. Yet, how different ENSO types contribute to these decadal variations and long-term variance trends remain uncertain. Here, we decompose the low-frequency changes of ENSO variance into contributions from ENSO diversity categories. We propose a fuzzy clustering of monthly SSTA to allow for non-binary event category memberships. Our approach identifies two La Niña and three El Niño categories and shows that the shift of ENSO variance in the mid-1970s is associated with an increasing likelihood of strong La Niña and extreme El Niño events.
\end{abstract}


\section{Introduction}

The El Niño-Southern Oscillation (ENSO), characterized by anomalous sea surface temperatures (SSTs) in the tropical Pacific, exhibits notable diversity in its amplitude, temporal evolution, and spatial pattern. The El Niño events of 1982-83 and 1997-98, for instance, recorded exceptionally high SST anomaly (SSTA) values in the eastern equatorial Pacific, whereas the El Niño of 2002-03 was less extreme and exhibited the largest anomalies in the central equatorial Pacific \citep{mcphaden2004}. In order to describe these event-to-event differences, El Niño events have been generally categorized as Eastern Pacific (EP), and Central Pacific (CP) types \citep{capotondi2015}. EP El Niño events typically have their peak SSTA in the eastern Pacific, may exhibit stronger intensities, and a largely reduced zonal thermocline slope, resulting in the pronounced discharge of warm water from the equatorial thermocline. In contrast, CP events show peak SSTA in the central Pacific and are comparatively weaker with smaller changes in zonal thermocline slope and warm water discharge \citep{kug2009,capotondi2013}.

These different types of ENSO events have substantially different downstream impacts on the global climate \citep{strnad2022,benichemargot2023}. For example, extreme drought conditions were recorded in eastern Australia in 2002, while a minor impact on precipitation was detected during the extreme 1997 event \citep{wang2007}. Weaker and shorter-lived CP events are associated with warm conditions in the equatorial Atlantic during boreal winter, while stronger and persistent EP events lead to cold anomalies in that area, with different impacts on precipitation over northeastern Brazil \citep{kao2009,rodrigues2011}. Thus, a deeper understanding of ENSO diversity is critical to support predictions of ENSO impacts.

ENSO characteristics, including amplitude and spatial pattern, exhibit decadal variations, which are mediated by changes in the background state of the tropical Pacific \citep{capotondi2023}. Notable decadal phase transitions were observed in the late 1970s \citep{miller1994} and around the year 2000 \citep{mcphaden2012}. Paleoclimate data also indicate an increase in ENSO amplitude over recent decades \citep{grothe2020}, consistent with modeling results showing a significant increase in ENSO amplitude after 1960, which was attributed to anthropogenic forcing \citep{cai2023}. Analysis of a large number of observationally-based datasets revealed that the longitudinal location of the maximum SST anomalies, as well as the intensity of both El Niño and La Nina events, undergo decadal fluctuations \citep{dieppois2021}, which can, in turn, modulate ENSO predictions \citep{lou2023}. However, event location and intensity were considered separately by \citet{dieppois2021}, so that the contribution of different ENSO types to these decadal changes and long-term trends in ENSO variance remains unclear. Paleoclimate data \citep{lawman2022} indicate that extreme ENSO events may contribute to increases in ENSO variance. Similarly, climate models that capture relevant aspects of ENSO nonlinearities project an increase in ENSO variance that is linked to an increase in the frequency of extreme ENSO events \citep{cai2021}. However, a more comprehensive understanding of these changes from an ENSO diversity perspective is still missing.

One main reason hindering the estimation of the contribution of different ENSO categories to its decadal modulation is that ENSO classifications often depend on the chosen definitions \citep{pascolini-campbell2015,yu2013,capotondi2020,abdelkaderdicarlo2023}. The disagreement between different ENSO classification methods is likely due to the assumption that ENSO events can be classified into binary types, based on indices capturing the location of the highest SSTA in the Tropical Pacific (like the Niño3 and Niño4 regions), or using Empirical Orthogonal Functions (EOFs) \citep{ashok2007, kug2009,kao2009,takahashi2011}. However, ENSO events are continuously distributed in the space spanned by the two leading principal components (PCs) \citep{takahashi2016,cai2018,capotondi2020}. Approaches that use a more continuous distribution of SSTAs to identify diversity show that events occur over multiple locations but with enhanced probabilities over the central and eastern tropical Pacific \citep{dieppois2021,shin2021}. In addition, both EP- and CP-type events appear to share common underlying dynamical processes, albeit with varying relevance depending on the longitude \citep{capotondi2013}.

A notable example of an El Niño that eludes a binary classification is the 2015/16 events. While its SSTA and several of its impacts were typical of extreme EP events \citep{santoso2017}, this event was not associated with a significant impact on California precipitation and was not followed by a strong La Niña, as other events of this type (i.e., 1982-83 and 1997-98). Indeed, this event was considered a mixture of EP and CP types \citep{paek2017,capotondi2020}.

Here, we propose a new approach for characterizing ENSO diversity to better understand its relationship with decadal changes in ENSO variance. To that end, we develop a fuzzy clustering of the low-dimensional representation of monthly SSTA in the PC1-PC2 space to achieve a non-binary event categorization (i.e., events belonging to one cluster or not). Instead, in this fuzzy clustering approach, individual events are assigned a probability of belonging to a given cluster. Such membership probabilities are then used to determine their relative contributions to the low-frequency ENSO variance.

\section{Methodology and data}
    \label{sec:methods}

\subsection{Data}

Our analysis is conducted on monthly SSTA in the tropical Pacific (130°E-70°W, 30°S-30°N) from eight reanalysis datasets merged together (Tab.~S1 and SI Sec.~S1). To ensure the same number of data points per month, we randomly select four data points per month from the eight datasets covering 1901--2022. While the dimensionality reduction encompasses all months, we select El Niño and La Niña winter months for the fuzzy clustering. Boreal winters (December-January-February (DJF)) are selected as El Niño (La Niña) when the average SSTA in the Niño3.4 region is larger than 0.5 K (smaller than -0.5 K).

\subsection{Fuzzy clustering}

Clustering algorithms are typically based on a distance measure between data points which becomes ill-posed in high-dimensional spaces \citep{parsons2004}. To mitigate this issue, we first reduce the dimensionality of the geospatial fields before applying fuzzy clustering using EOF analysis.

Our input data, $X = \left( x(t_{1}),\ldots,x( t_{N}) \right)$, which consists of tropical Pacific SSTA fields $x(t) \in \cal{R}^{N_{\rm{lon}} \cdot N_{\rm{lat}}}$, are projected onto the first two EOF patterns to obtain the two PCs $z(t) \in \cal{R}^{2}$, referred to as a latent vector. Besides EOFs, we also use a nonlinear dimensionality reduction method, specifically a convolutional autoencoder neural network (SI Sec.~S2, Fig.~S1).

We apply the Gaussian Mixture Model (GMM), a probabilistic unsupervised clustering approach to identify different types of ENSO events. GMMs describe each cluster via a multivariate Gaussian distribution, accommodating overlapping probabilities. Mathematically, GMMs assume that the probability distribution of latent states, $p(z)$, comprises a mixture of Gaussians:
\begin{equation}
    p(z) = \sum_{k=1}^K \pi_{k} \mathcal{N}\left( z \mid \mu_{k},\Sigma_{k} \right),
    \label{eq:gmm}
\end{equation}
with $K$ representing the number of Gaussians, $\mathcal{N}\left( z \mid \mu_{k},\Sigma_{k} \right)$, each characterized by a mean $\mu_{k}$ and covariance $\Sigma_{k}$. The probability of each Gaussian is denoted as $p\left( c_{k} \right) = \pi_{k}$ with $\sum_{k=1}^K \pi_{k} = 1$. The optimal number of clusters $K$ is obtained, by minimizing the Bayesian Information Criterion (BIC). The BIC balances model complexity and fit quality, as described by \citet{schwarz1978} and in SI Sec.~S3.

\begin{figure}[!b]
    \includegraphics[width=1.0\textwidth]{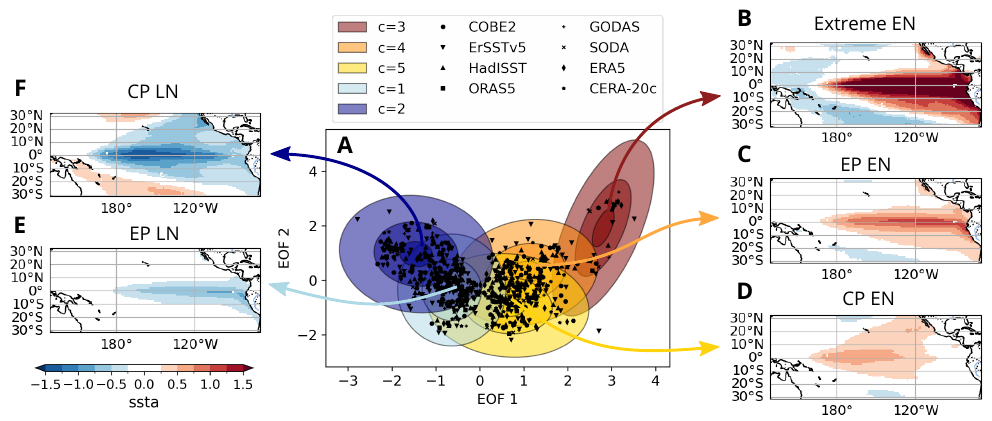}
    \caption{\textbf{El Niño and La Niña categories in PC1-PC2 space.} Monthly boreal winter (DJF) SSTAs of El Niño and La Niña events of all reanalysis products (see SI Tab. S1) projected onto the PC1-PC2 space and fitted by a Gaussian Mixture Model (GMM). Each event (DJF averages are shown as black dots in A) has a probability of belonging to each of the five categories (colored Gaussians in A). Pacific SSTA composites for each category are obtained by using the category membership as weights for the averages, depicted in panels (B-F). We obtain three El Niño-like patterns: \emph{Extreme EN} (B), \emph{Strong EN} (C), and \emph{Weak EN} (D), while La Niña events form two categories: \emph{Weak LN} (E) and \emph{Strong LN} (F).}
    \label{fig:pcgmm-latent}
\end{figure}

\subsection{Variability estimation}
The GMM allows assigning a probability, $p\left(c_{k}|z(t) \right)$, to each data point, $z(t)$, that quantifies its likelihood of belonging to category $c_k$. These are the category memberships of the data point (Sec.~S4), which are inherently fuzzy and non-binary (i.e. they are probabilities between 0 and 1), and allow us to model ENSO events in terms of their likelihood of occurrence. As a consequence, we can decompose a variable $y(t)$, for instance, SSHA, into the contributions of each category, by
\begin{equation}
    y(t) = \sum_k p\left( c_{k}|z(t) \right) \cdot y(t) : = \sum_k y_{k}(t)
    \label{eq:temporal_variability}
\end{equation}
where $y_{k}(t)$ is defined as the contribution of category $c_{k}$. Averaging $y_k(t)$ over time corresponds to a weighted average, with the weights being the categorical memberships (Sec.~S5). Similarly, we can write the variance $\langle y^2 \rangle$, as
\begin{equation}
    \langle y^{2}\rangle = \langle \left(\sum_k y_{k} \right)^2 \rangle = \sum_k \langle y_{k}^2\rangle + \sum_{l\neq m}\langle y_{l},y_{m}\rangle,
    \label{eq:temporal_variability_contribution}
\end{equation}
where $\langle y_{k}^{2}\rangle$ is the variance contribution of category $c_k$ and $\langle y_{l},y_{m}\rangle$ is the co-variability of categories $l$ and $m$.

\section{Results} \label{sec:results}

\subsection{ENSO diversity is well explained by five categories} \label{sec:gmm_catergories}

The two most dominant EOFs of all eight SSTA datasets combined (Tab.~S1) present the well-known spatial patterns associated with ENSO, i.e. EOF1 depicts the typical ENSO pattern with anomalies in the Central-Eastern Pacific, while EOF2 exhibits an east-west dipole structure (Fig.~S2). Projection of the monthly SSTA of all boreal winter (December-January-February) El Niño and La Niña events onto EOF1 and EOF2, produces a distribution in the corresponding PC1–PC2 space that exhibits a wide, boomerang-like shape (Fig.~\ref{fig:pcgmm-latent}A, Fig.~S3). 
This nonlinear relationship between PC1 and PC2 has been considered an expression of key ENSO dynamics \citep{cai2018,takahashi2016,ham2012,karamperidou2017}, and used in the selection of models to consider for examining future projections \citep{cai2018,cai2021}. The nonlinear relationship itself can be accounted for by using a nonlinear transformation in place of the linear EOF-based transform, e.g., an autoencoder, and in that case, the boomerang-like shape is replaced by a simple linear relation between the two latent dimensions (SI Fig.~S4 and SI Sec.~S2).

The distribution of boreal winter El Niño and La Niña months in the PC1-PC2 space do not exhibit clear gaps visually, however, their density varies, suggesting a categorical structure \citep{takahashi2016}. We model this distribution using a GMM with $k = 5$ categories (Eq.~\ref{eq:gmm}), a number determined using the BIC to ensure parsimony (see SI Sec.~S3, Fig.~S5A). The five Gaussians are arranged in a row along the boomerang-shaped distribution from the coldest events at the leftmost tip to the warmest events at the rightmost end (Fig.~\ref{fig:pcgmm-latent}A).

\begin{figure}[!b]
    \centering
    \includegraphics[width=1.0\textwidth]{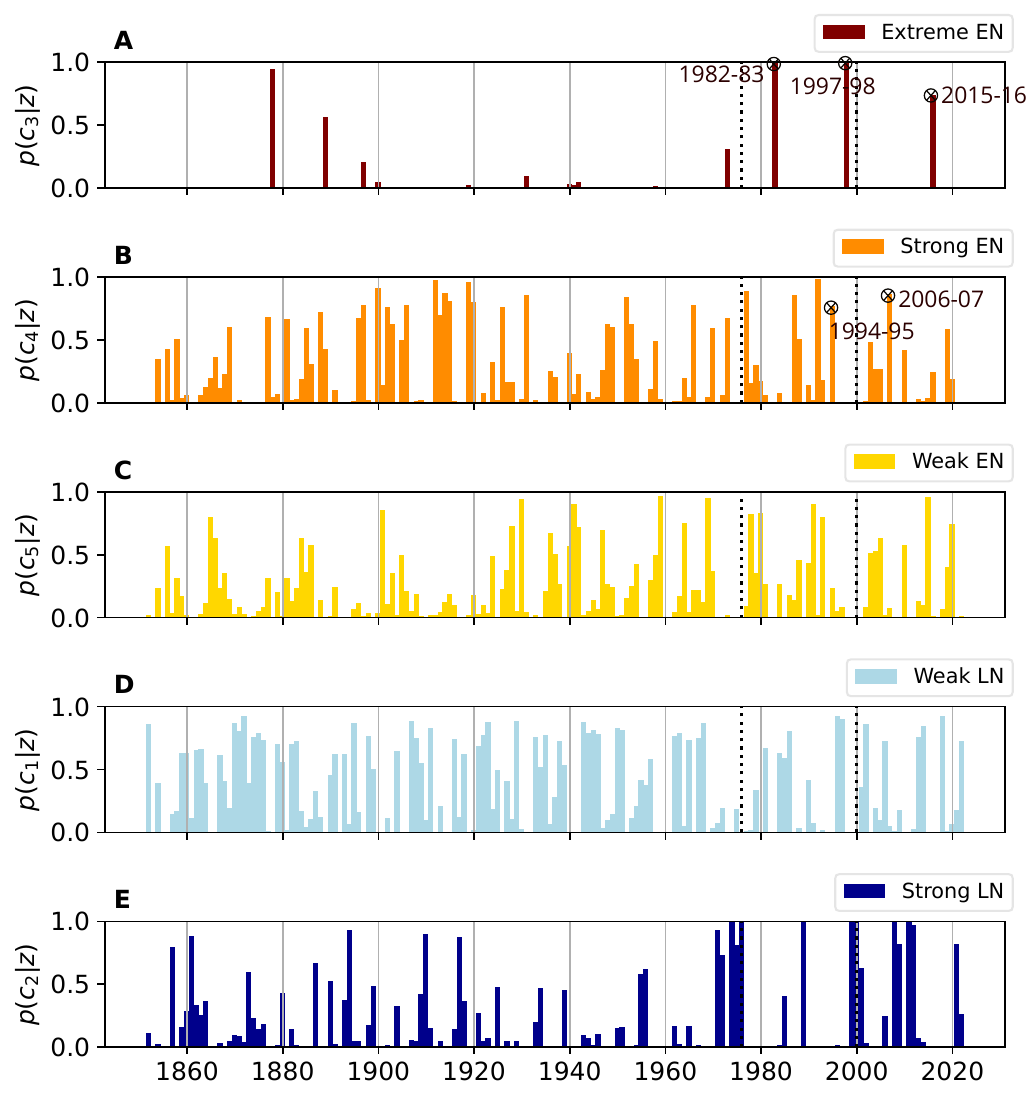}
    \caption{\textbf{Probabilistic category membership.} The GMM in Fig.~\ref{fig:pcgmm-latent} allows us to estimate the likelihood, $p\left( c_{k}|z(t) \right)$, of each El Niño and La Niña winter month, $z(t)$, to belong to each of the categories, $c_{k}$ (SI Sec.~S4). An event belongs only to one category when its probability is 1. However, many events have shared probabilities across several categories. The categories are sorted in the following order (top to bottom): \emph{Extreme EN} (A), \emph{Strong EN} (B), \emph{Weak EN} (C), \emph{Weak LN} (D), and \emph{Strong LN} (E). For visual reasons, we average the monthly probabilities over each winter (DJF) and over reanalysis products. The dashed lines indicate the reported shifts in ENSO variability in 1976-77 and 2000.}
    \label{fig:pcgmm-weights}
\end{figure}

We use the category memberships, $p\left( c_{k}|z(t) \right)$, as weights for averaging the Pacific SSTAs of each category (SI Sec.~S5) and find three El Niño-like patterns (Fig.~\ref{fig:pcgmm-latent}B-D), and two La Niña-like patterns (Fig.~\ref{fig:pcgmm-latent}E-F). Besides the different zonal locations of maximum warming/cooling, their defining factor is the tropical Pacific SSTA intensity. Hence, we will refer to the three El Niño categories as \emph{Extreme EN} (Fig.~\ref{fig:pcgmm-latent}B), \emph{Strong EN} (Fig.~\ref{fig:pcgmm-latent}C) and \emph{Weak EN} (Fig.~\ref{fig:pcgmm-latent}D), and correspondingly to \emph{Weak LN} (Fig.~\ref{fig:pcgmm-latent}E) and \emph{Strong LN} (Fig.~\ref{fig:pcgmm-latent}F) for the La Niña categories. We find that the overall clustering of the SSTA patterns into the five categories is robust (SI Sec.~S6) upon changing the number of EOFs (SI Fig.~S5B, Fig.~S6, Fig.~S7B), incorporating SSHA data along with the SSTA (SI Fig.~S8 K-O), or varying the latitudinal range of the input (SI Fig.~S8F-J). For comparison, we show the Gaussians for k=4 (SI Fig.~S9), and k=6 (SI Fig.~S10) categories.

The mean category membership for each boreal winter averaged across the datasets results in five time series of probabilities --- one for each category --- which reflect the likelihood of their occurrences (Fig.~\ref{fig:pcgmm-weights}). We find that the indices are not sensitive to the dataset used for their estimation, as seen in the relatively small spread of the membership probabilities over the datasets at each time point (SI Fig.~S11). As with the number of clusters, the category membership is also robust to changes in the number of variables, spatial domain, and the use of nonlinear encoding (SI Fig.~S12). 

Comparing the category membership series for \emph{Strong} and \emph{Weak EN} events (Fig.~\ref{fig:pcgmm-weights}B, C) to two conventional classifications -- the Niño3-Niño4 classification by \citet{kug2009} and the EOF-based E/C classification by \citet{takahashi2011} -- we find that our classification mostly agrees with them (SI Tab.~S2, Sec.~S7).  Differences, when they do occur, correspond to differences between the other classifications as well \citep{pascolini-campbell2015,yu2013,capotondi2015}. The two El Niños of 1994-95 and 2006-07, however, are classified by both conventional classifications as CP events, whereas we find them to be \emph{Strong EN} events (Tab.~S2). This is likely because the warm water anomalies span from the Central to the Eastern Pacific during the duration of these events, leading to possible ambiguities in the event definition (SI Fig.~S13C, H). We also note that unlike the 1982-83 and 1997-98 events, which unambiguously belonged to the \emph{Extreme EN} category, the 2015-16 event also has a nonzero membership in the \emph{Strong EN} category, confirming its mixed nature.

\subsection{Extreme EN events are different from Strong EN events}
    \label{sec:extreme-en-events-are-different-from-strong-en-events}

\begin{figure}[!t]
    \centering
    \includegraphics[width=1.0\textwidth]{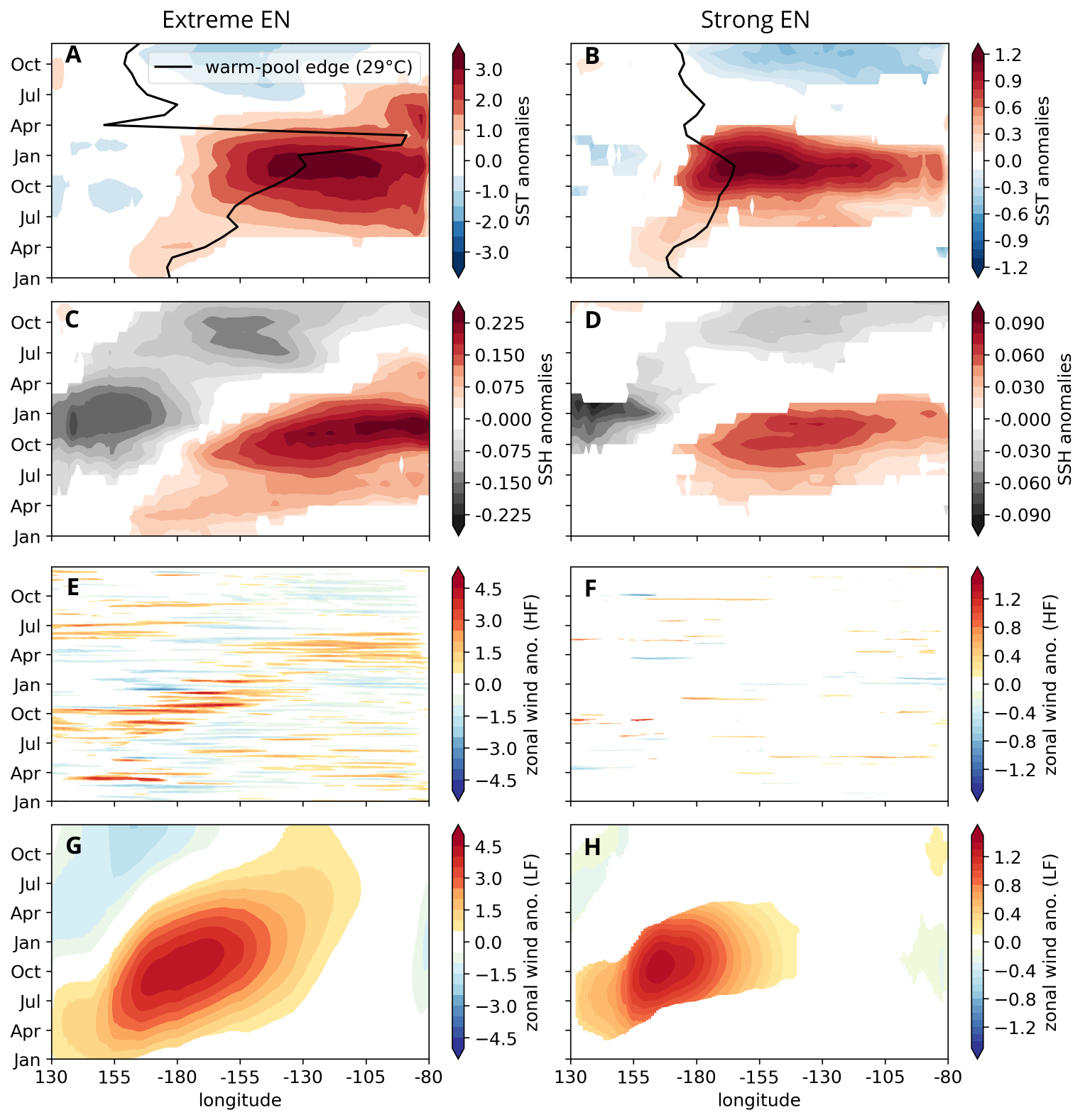}
    \caption{ \textbf{\emph{Extreme EN} and \emph{Strong EN} category.} Hovmöller diagrams of SSTA (A, B), SSHA (C, D), high-frequency (HF) zonal wind anomalies (E, F), and low-frequency (LF) zonal wind (G, H) anomalies are obtained by meridional averages (5°S - 5°N) of each month in the year preceding and succeeding El Niño events. Each two-year period is weighted by the corresponding DJF average category membership probability (Fig.~\ref{fig:pcgmm-weights}). The black line in (A) and (B) indicates the warm-pool edge, i.e., the 29°C SST isotherm (SI Sec.~S8). Only values that are statistically significant above the 95th percentile are displayed (SI Sec.~S9). SSTA and SSHA are taken from ORAS5 (1958--2022), while 10-meter zonal winds, with their HF- and LF components (Sec.~S1) are computed from ERA5.}
    \label{fig:en_hovmoeller}
\end{figure}

A key distinction of our ENSO categorization from conventional methods is the identification of the extreme El Niños as a separate class. While our \emph{Weak EN} category corresponds to the conventional CP El Niño type, the conventional EP El Niño type is split into \emph{Extreme EN} and \emph{Strong EN} categories (SI Sec.~S7, Tab.~S2). SSTA composites of conventional EP El Niño events (SI Fig.~S14A, C, E, Fig.~S15B) exhibit the maximum warming in the eastern Pacific. The maximum warming is however strongly influenced by the few extreme EN events (1982/83, 1997/98, 2015/16), with an eastward shift of the peak towards the central Pacific when we exclude the extreme events (SI Fig.~S14B, D, F).

The separation of \emph{Extreme EN} from \emph{Strong EN} events highlights differences in their impacts. We find that during \emph{Extreme EN} events, statistically significant (at 95 \% confidence) warmer-than-average SSTs and negative OLR anomalies occur in the Indian Oceans (SI Fig.~S16A, K), which are not significant for \emph{Strong EN} (SI Fig.~S16B, L).

The two categories also differ in their evolution, analyzed using Hovmöller diagrams for the year preceding and succeeding an event (Fig.~\ref{fig:en_hovmoeller}, SI Fig.~S17). \emph{Extreme EN} events show significant warm water volume anomalies, as described by SSHA, around the dateline already in the spring before an event, corresponding with a shift of the warm pool edge near the dateline (black line Fig.~\ref{fig:en_hovmoeller}A, C). The \emph{Extreme EN} onset phase also shows strong positive HF and LF zonal wind components to the west of the dateline, during the preceding spring, extending further east as the event develops to its mature phase (Fig.~\ref{fig:en_hovmoeller}E, G). \emph{Strong EN} events, on the other hand, do not demonstrate a consistent onset pattern in the preceding spring (Fig.~\ref{fig:en_hovmoeller}D, F, H). 

These findings corroborate ideas presented in prior research on the impact of the Walker circulation's zonal shift \citep{thual2023} and the influence of stochastic high-frequency winds, called Westerly Wind Bursts (WWBs), on ENSO Diversity \citep{fedorov2015,capotondi2018,puy2019}.

\subsection{Interdecadal ENSO variability is driven by Strong and Extreme events}
\label{sec:interdecadal_variability}

\begin{figure}[!t]
    \centering
    \includegraphics[width=1.0\textwidth]{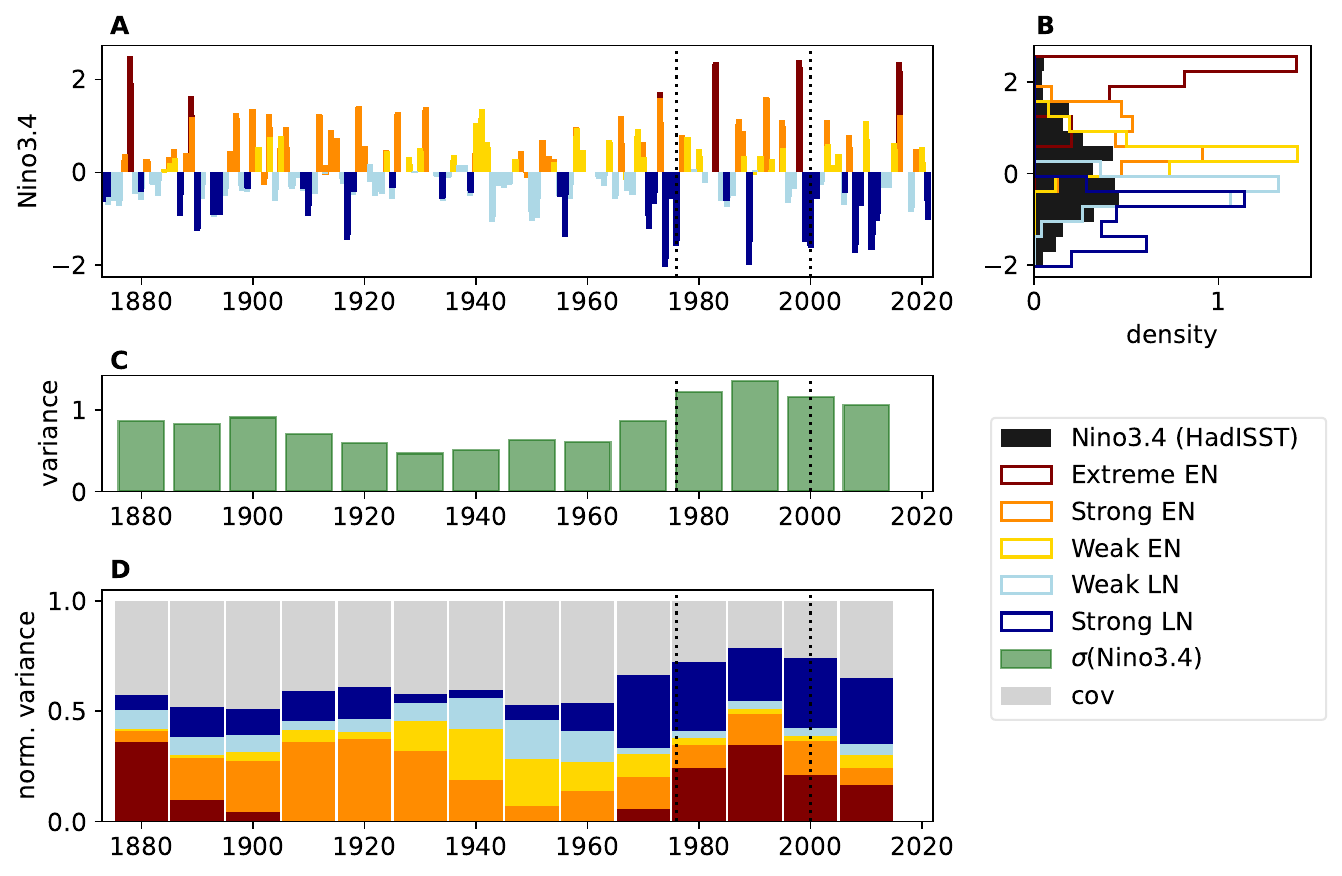}
    \caption{ \textbf{Low-frequency changes of ENSO variance.} For each category, the Niño3.4 index is multiplied by their category membership probabilities (A). The histogram of Niño3.4 intensities (B), highlights different SSTA amplitudes between categories. The 20-year running variance of Niño3.4 every 10 years (C), is used to normalize the 20-year running variance of each category (D). \emph{Extreme EN} and \emph{Strong EN} categories dominate the Niño3.4 variance in the early and late 20th century. The variance shift in 1976-77 and 2000 (dashed lines) highlight reported changes in ENSO variability. The Niño3.4 index is taken from HadISST \citep{rayner2003}.}
    \label{fig:decadal_variability}
\end{figure}

The membership probabilities of each category (Fig.~\ref{fig:pcgmm-weights}A, B) encode a distinct pattern of decadal-to-multidecadal ENSO variability. In particular, the \emph{Strong LN} and the \emph{Extreme EN} categories show a markedly prominent low-frequency variability, which is less evident in other categories. To quantify the low-frequency variation of ENSO, we multiply the Niño3.4 index (from HadISST) by the corresponding membership probabilities for each category (Eq.~\ref{eq:temporal_variability}, Fig.~\ref{fig:decadal_variability}A) and calculate their 20-year running variances every 10 years (Eq.~\ref{eq:temporal_variability_contribution}). The variance of each category is normalized by the 20-year running variance of the Niño3.4 index (Fig.~\ref{fig:decadal_variability}C) to determine their relative contributions. 

The total variance of the Niño3.4 index shows a low-frequency modulation, with a minimum around 1920-1960, followed by an increasing trend, which is consistent with the shift in the mean and variance of ENSO after the so-called “1976-77 climate shift” \cite<e.g.>[]{miller1994} (dashed line in Fig.~\ref{fig:decadal_variability}C). A slight decrease in variance is noticeable after the year 2000, in line with the other reported climate shift \cite<e.g.>[]{mcphaden2012}. While changes are noticeable in the variance of all categories, only minor contributions to the total Niño3.4 variance are observed for the \emph{Weak LN} and \emph{Weak EN} categories (Fig.~\ref{fig:decadal_variability}D). The \emph{Strong LN} category exhibits a substantial contribution starting around 1970, and the \emph{Strong EN} category dominates the variance from 1890 to $\sim$1930. Meanwhile, the contribution of the \emph{Extreme EN} category is particularly visible towards the end of the 19th century and after 1980. These results are robust against variations in window size (SI Fig.~S18) and using different reanalysis products (SI Fig.~S19). 

Between 20-40\% of the decadal variance is due to the covariance between the different ENSO categories (denoted as \emph{cov} in Fig.~\ref{fig:decadal_variability}D). The covariance captures mainly the contribution of co-occurrences of different categories to the overall variance in a decade. However, other factors such as events with intermediate values of category memberships and statistical errors in estimating the GMM categories also influence the covariance. 
Our statistical approach does not allow us to assess the dynamics underpinning these decadal changes in the contribution of the different ENSO categories. ENSO decadal modulation is often associated with changes in the mean state, but it is not clear whether mean state changes are induced by influences from the extratropical Pacific or other ocean basins, or result as a residual of random variations of ENSO itself \citep{capotondi2023}. Some studies, however, suggest that some ENSO events could induce decadal phase transitions through either nonlinear dynamical heating \citep{liu2022} or by inducing a discharge of upper ocean heat content in the off-equatorial Western Pacific \citep{meehl2021}.

Our results provide novel insights into low-frequency changes in ENSO variance, which combines changes in frequency and intensity. We find that although the magnitude of total variance contributed by Eastern Pacific warming events (i.e. \emph{Extreme EN} and \emph{Strong EN events} combined) since the turn of the 20th century is comparable to that seen during the period from the 1880s to 1940s, the variance in the recent decades has been dominated by the \emph{Extreme EN} category, a result consistent with the statistically significant change in ENSO properties in the late 1970s \citep{capotondi2017}. This shift was associated with a weakening of the easterly trade winds and a zonal reduction of the equatorial thermocline slope, conditions favoring stronger El Niño events. Linked to the increased contributions from \emph{Extreme EN events} is the concurrent increase in the contributions of \emph{Strong LNs} since the 1970s (dark blue in Fig.~\ref{fig:decadal_variability}D), manifested in very strong, multi-year events starting around 1970 (Fig.~\ref{fig:decadal_variability}A). These results are consistent with a higher likelihood of stronger CP La Niñas following the heat discharge of strong El Niños, as exemplified by the 1998-99 La Niña event after the extreme 1997-98 El Niño \citep{cai2015, geng2023}. While the increasing contribution of \emph{Extreme EN} and \emph{Strong LN} to ENSO variance starting around 1970 aligns with results from climate model simulations \citep{cai2021,gan2023}, our findings more specifically highlight the key role played by \emph{Strong LN}  in the ENSO variance changes after 1970.

We also find that during the `quiescent’ period of ENSO, roughly from the 1930s to the 1960s, most of the ENSO variability originated from Central Pacific warming and Eastern Pacific cooling events, i.e., from the weaker event types, while the influence of Central Pacific warming on the ENSO variance in recent decades has been almost negligible. This latter result is in apparent disagreement with the reported increase in intensity and frequency of CP events since 1980 \citep{lee2010}, and especially after 2000 (dashed line in Fig.~\ref{fig:decadal_variability}D; \citep{mcphaden2011}). This discrepancy may be related to the 20-year running variance we used to construct Fig.~\ref{fig:decadal_variability}D, and our inclusion of the 2015/16 El Nino, an event that was missing in the records used by the earlier studies.

\section{Discussion} \label{sec:discussion}

We present a new approach for studying the influence of ENSO diversity on low-frequency changes in ENSO variance. In particular, we use a Gaussian Mixture Model (GMM) within the low-dimensional PC1-PC2 space of monthly SSTA, which enables the assignment of non-binary event category memberships. We identify two La Niña categories (\emph{Weak LN}, \emph{Strong LN}) and three El Niño categories (\emph{Extreme EN, Strong EN, Weak EN}), which combine the two dimensions of ENSO diversity --- longitudinal location of maximum SSTA and its intensity. A key contribution of our work involves utilizing the membership probabilities to determine how each of the five categories individually affects the overall decadal variability in the Niño3.4 region. We find that the increasing frequency of both \emph{Extreme EN} and \emph{Strong LN} are the primary drivers of the increased ENSO variance post-1970. While these findings are consistent with previous studies that also detected an increase in extreme ENSO events in the second half of the 20th century \citep{cai2018,cai2023}, our results further highlight the key role played by the increasing frequency of \emph{Strong LN} in these ENSO variance changes. 

The proposed fuzzy clustering approach could also be used to quantify how well climate models represent ENSO diversity, akin to \citet{dieppois2021} and \citet{ayar2023}. A preliminary analysis of the Community Earth System Model version 2 (CESM2) shows that this model only exhibits four ENSO categories instead of five (SI Sec.~S10, Fig.~S20), a discrepancy which cannot be explained by the different number of samples or other technical choices. Specifically, the \emph{Extreme EN} category is missing in the model. While ENSO in CESM2 has an amplitude larger than observed, modeled SSTAs tend to occur preferentially in the central equatorial Pacific, with a more limited range of ENSO spatial patterns \citep{capotondi2020a}, a behavior that seems to align with the model’s inability to simulate extreme events in the eastern Pacific. In addition, the model appears to underestimate ENSO’s nonlinearities, as quantified by the quadratic fit coefficient, $\alpha$, in PC1-PC2 space  \citep{dommenget2013,karamperidou2017,cai2018}, which is smaller in CESM2 compared to reanalysis (Fig.~S21). An extensive examination of several climate models using our methodology is beyond the scope of this paper. However, our analysis of CESM2 demonstrates the potential value of our approach for achieving a detailed assessment of ENSO diversity in climate models. Such an investigation, as well as the analysis of possible changes in ENSO types in future climate scenarios, will be considered in subsequent studies.

\section*{Open Research}
The data on which this article is based are publicly available in \citet{zhang2019,hersbach2020,cobe2006,rayner2003,oras52021,giese2011,behringer1998,laloyaux2018}, with their details are listed in SI Tab.~S1. Our code is publicly available under \citet{schlor2023} and \emph{https://github.com/jakob-schloer/LatentGMM.git}.

\section*{Acknowledgments}
The authors thank the International Max Planck Research School for Intelligent Systems (IMPRS-IS) for supporting J. Schlör and F. Strnad. Furthermore, we express our gratitude to the NOAA Physical Science Laboratory for making their resources available for this study. Funded by the Deutsche Forschungsgemeinschaft (DFG, German Research Foundation) under Germany’s Excellence Strategy – EXC number 2064/1 – Project number 390727645. We acknowledge support from the Open Access Publication Fund of the University of Tübingen. A. Capotondi was supported by the NOAA Climate Program Office’s Climate Variability and Predictability (CVP) and Modeling, Analysis, Predictions and Projections (MAPP) programs and by DOE Award No. DE-SC0023228.

% References
\newpage
\bibliographystyle{unsrtnat}
\bibliography{references}

\end{document}